\documentclass[12pt]{article}
\usepackage[latin1]{inputenc}

\usepackage{amsmath}
\usepackage{color}
\usepackage{amsfonts}
\usepackage{amssymb}
\usepackage{graphicx}
\usepackage{geometry}
\usepackage{amssymb,epsfig,subfigure}
\usepackage{hyperref}
\usepackage{comment}
\usepackage{tabularx}
\usepackage{bm}
\usepackage{euscript}
\usepackage{graphicx}
\usepackage{color}
\usepackage{amsfonts}
\usepackage{exscale}
\usepackage{amsbsy}
\usepackage{subfigure}
\usepackage{textcomp}
\usepackage{comment}
\usepackage{hyperref}
\usepackage{slashed}
\usepackage{authblk}

\usepackage{tabularx}
\usepackage{euscript}
\usepackage{graphicx}
\usepackage{color}
\usepackage{amsfonts}
\usepackage{exscale}
\usepackage{amsbsy}
\usepackage{subfigure}
\usepackage{textcomp}
\usepackage{comment}
\usepackage{hyperref}
\usepackage{bm}

\def\be{\begin{equation}}
\def\ee{\end{equation}}
\def\bea{\begin{eqnarray}}
\def\eea{\end{eqnarray}}
\def\ba#1\ea{\begin{align}#1\end{align}}
\def\bg#1\eg{\begin{gather}#1\end{gather}}
\def\bm#1\em{\begin{multline}#1\end{multline}}
\def\bmd#1\emd{\begin{multlined}#1\end{multlined}}

\usepackage[numbers,sort&compress]{natbib}

\def\simge{
    \mathrel{\rlap{\raise 0.511ex 
        \hbox{$>$}}{\lower 0.511ex \hbox{$\sim$}}}}

\def\simle{
    \mathrel{\rlap{\raise 0.511ex 
        \hbox{$<$}}{\lower 0.511ex \hbox{$\sim$}}}}

\makeatletter
\renewcommand\section{\@startsection {section}{1}{\z@}%
                                 {-3.5ex \@plus -1ex \@minus -.2ex}
                                   {2.3ex \@plus.2ex}%
                                   {\normalfont\large\bfseries}}
\renewcommand\subsection{\@startsection{subsection}{2}{\z@}%
                                   {-3.25ex\@plus -1ex \@minus -.2ex}%
                                     {1.5ex \@plus .2ex}%
                                     {\normalfont\bfseries}}
\renewcommand\subsubsection{\@startsection{subsubsection}{3}{\z@}%
                                   {-3.25ex\@plus -1ex \@minus -.2ex}%
                                     {1.5ex \@plus .2ex}%
                                     {\normalfont\itshape}}
\makeatother

\def\pplogo{\vbox{\kern-\headheight\kern -29pt
\halign{##&##\hfil\cr&{\ppnumber}\cr\rule{0pt}{2.5ex}&\ppdate\cr}}}
\makeatletter
\def\ps@firstpage{\ps@empty \def\@oddhead{\hss\pplogo}%
  \let\@evenhead\@oddhead 
}
\def\maketitle{\par
 \begingroup
 \def\thefootnote{\fnsymbol{footnote}}
 \def\@makefnmark{\hbox{$^{\@thefnmark}$\hss}}
 \if@twocolumn
 \twocolumn[\@maketitle]
 \else \newpage
 \global\@topnum\z@ \@maketitle \fi\thispagestyle{firstpage}\@thanks
 \endgroup
 \setcounter{footnote}{0}
 \let\maketitle\relax
 \let\@maketitle\relax
 \gdef\@thanks{}\gdef\@author{}\gdef\@title{}\let\thanks\relax}
\makeatother

\numberwithin{equation}{section}

\begin{document}

\setcounter{page}0
\def\ppnumber{\vbox{\baselineskip14pt
}}

\def\ppdate{
} \date{\today}

\title{\bf Two-dimensional conductors with interactions and disorder from particle-vortex duality
\vskip 0.5cm}
\author[1]{H. Goldman}
\author[2]{M. Mulligan}
\author[3,4]{S. Raghu}
\author[5]{G. Torroba}
\author[3]{M. Zimet}
\affil[1]{\small \it Department of Physics and Institute for Condensed Matter Theory,\protect\\\small \it University of Illinois at Urbana-Champaign, Urbana, IL 61801, USA}
\affil[2]{\small \it Department of Physics and Astronomy, University of California,
Riverside, CA 92511, USA}
\affil[3]{\small \it Department of Physics, Stanford University, Stanford, CA 94305, USA}
\affil[4]{\small \it SLAC National Accelerator Laboratory, 2575 Sand Hill Road, Menlo Park, CA 94025, USA}
\affil[5]{\small \it Centro At\'omico Bariloche and CONICET, Bariloche, Rio Negro R8402AGP, ARG}

\bigskip

\maketitle

\begin{abstract}
We study Dirac fermions in two spatial dimensions (2D) coupled to strongly fluctuating U(1) gauge fields in the presence of quenched disorder.  
Such systems are dual to theories of free Dirac fermions, which are vortices of the original theory. In analogy to superconductivity, when these fermionic vortices localize, the original system becomes a perfect conductor, and when the vortices possess a finite conductivity, the original fermions do as well.  
We provide several realizations of this principle and thereby introduce new examples of strongly interacting 2D metals that evade Anderson localization.
 \end{abstract}
\bigskip

\newpage

\tableofcontents

\vskip 1cm

\section{Introduction}\label{sec:intro}

With very few exceptions,\footnote{We neglect spin-orbit coupling in this paper \cite{Hikami80}.} a non-interacting disordered electron gas undergoes Anderson localization \cite{Wegner1976,Wegner1979,Abrahams1979,Hikami1981,Mirlin2008,2010IJMPB..24.1526W} in two spatial dimensions (2D).   
A fundamental and open question has been whether strong interactions can alter this conclusion \cite{Castellani1984,Chakravarty1998,Castellani1998,Punnoose2005,DasSarma2014}.  
From the experimental standpoint, behavior suggestive of a genuine zero-temperature ($T=0$) metallic phase has been observed in several two dimensional systems \cite{Jaeger1989,Kravchenko1995,Ephron1996,Wong1996,Mason1999,Kravchenko2003,Qin2006, Zudov2015, Tsen:2016aa, Breznay2017, 2017arXiv170801908W}.  
However, because experiments are always conducted at $T > 0$, these experimental findings have often been dismissed as finite temperature crossovers. 
Instead, the current belief is that  2D metallic ground states generically do not exist, except at isolated quantum critical points \cite{Fisher1990}.  
Therefore, the existence of unambiguous theoretical examples of 2D metallic phases would change our current understanding of disordered, interacting electron systems.  
The goal of this paper is to provide several such examples. 

Since 2D Fermi liquids are unstable to disorder \cite{Altshuler1980, Castellani1998}, it seems reasonable to look for metallic phases in a system without quasiparticles \cite{Chakravarty1998}.   
From effective field theory considerations, there is a well-known way of destroying  Fermi liquid  quasiparticles: by coupling them to gapless bosons.  In the condensed matter setting, the choice of gapless bosons includes order parameter fluctuations or gauge fields.
The former requires tuning to a quantum critical point and hence can never describe a zero-temperature metallic {\it phase}.\footnote{This is because away from the critical point, Fermi liquid behavior recurs at length scales large compared to the order parameter correlation length.}  
By contrast, gauge invariance ensures that gauge fields remain gapless without fine tuning. Therefore, we are naturally led to ask whether 2D metals coupled to fluctuating gauge fields may be stable to disorder.  
The familiar example of the  electromagnetic gauge field destroys metallic quasiparticles only at extremely low energy scales \cite{Holstein1973}.  
The situation can be quite different with {\it emergent} gauge fields, which
naturally arise in dual constructions of strongly coupled 2D systems.  
Moreover, in several of the experimentally observed metallic phases, emergent gauge fields are believed to play a crucial role in the low-energy dynamics.  
Thus, the study of fermions coupled to gauge fields with disorder is motivated both from theoretical and phenomenological considerations.  

In this paper, we study the effects of disorder on some of the simplest such gauge theories, which consist of Dirac fermions coupled to strongly fluctuating $U(1)$ gauge fields.  
We generically refer to such theories as QED$_3$.\footnote{QED$_3$ stands for quantum electrodynamics (or more precisely, an abelian gauge theory coupled to Dirac fermions) in $2+1$ spacetime dimensions.} 
Much work has studied the effects of disorder on Dirac fermion systems without gauge fields \cite{Fradkin1986,Fradkin1986a, Ludwig1994,Denis2002,Ryu2007, Foster2008, PhysRevB.95.075131}.  
However, gauge fields play a crucial role here in mediating singular interactions among the fermions.  
Furthermore, unlike electromagnetism, the  gauge fields studied here propagate in $2+1$ spacetime dimensions.  
Simple scaling arguments for such a system reveal that the coupling of matter to gauge fields is strongly relevant in the renormalization group (RG) sense.  
Thus, at low energies, the system flows toward strong coupling, rendering the study of disorder effects highly challenging.  
As a consequence, one has two options: 
to study the problem in a controlled perturbative framework, for instance, by studying various large N limits, or to search for non-perturbative solutions.  

We consider disorder effects in the strong coupling limit of QED$_3$ using duality arguments.
We study the case where the Fermi level is at the Dirac point, which leads to a vanishing density of states in the clean limit. In a follow-up publication, we will consider the case when the clean limit admits Fermi surfaces of Dirac fermions.  

\section{Basic setup}\label{sec:setup}

The defining property of a metal is a non-zero dc conductivity,
\begin{equation}
\sigma^{\rm dc} = \lim_{T \rightarrow 0}
 \lim_{\omega \rightarrow 0} 
\sigma_{xx}\left(\omega, T \right),
\end{equation}
where $\sigma_{xx}(\omega, T)$ is the finite-temperature ac longitudinal conductivity.
Note that this definition can accommodate systems with a vanishing density of states (that are incompressible).
A familiar example  is clean graphene with $1/r$ Coulomb interactions.  
Such a system has a finite $\sigma^{\rm dc}$ but is unstable to weak potential disorder, which acts as a marginally relevant perturbation.  
In this case, the system develops a finite density of states at low energies and eventually undergoes Anderson localization.  
We instead study Dirac fermions coupled to a $2+1$ dimensional gauge field (which mediates logarithmic, rather than $1/r$ interactions).  

Our starting point is the following low-energy effective Lagrangian in 2+1 spacetime dimensions:
\begin{align}
L_{\mathrm{QED}_3}  = i \bar \psi \slashed D_a \psi - \frac{1}{4g^2} f_{\mu \nu}^2, 
\end{align}
where $\mu \in \{t,x,y \}$, $\slashed D_a = \gamma^\mu (\partial_\mu - i a_\mu)$, $\gamma^\mu = (\sigma^3, i \sigma^1, i \sigma^2)$ where $\sigma^j$ are Pauli matrices,\footnote{This choice corresponds to a metric with signature $\eta_{\mu\nu}=\text{diag}(+,-,-)$.} $\psi$ is a 2-component Dirac fermion for which $\bar{\psi} = \psi^\dagger \gamma^t$, and $a_{\mu}$ is a dynamical $U(1)$ gauge field with gauge charge $g$ (the analog of the electric charge $e$ in ordinary electromagnetism) and field strength $f_{\mu \nu} = \partial_{\mu} a_{\nu} - \partial_{\nu} a_{\mu}$.
In the absence of disorder, $L_{\rm QED_3}$ preserves charge conjugation symmetry ${\bf C}$, but breaks parity ${\bf P}$ (reflection along one spatial direction), time-reversal ${\bf T}$, and particle-hole symmetry ${\bf PH}={\bf CT}$.
These symmetries can be restored if we consider an even number of Dirac fermions.

We perturb $L_{\mathrm{QED}_3}$ with the following types of disorder:
\begin{eqnarray}
V_{\vec{A}} &=& \bar \psi \vec \gamma \cdot \vec{A}(\vec{x}) \psi,  \nonumber \\
V_M &=& M(\vec{x}) \bar \psi \psi, \nonumber \\
V_U &=& U(\vec{x}) \psi^{\dagger} \psi,
\end{eqnarray}
which correspond to flux, mass, and potential disorder.  
In each case, the disorder potential, $\vec{A}(\vec{x})$, $M(\vec{x})$, or $U(\vec{x})$,  only varies in space and is chosen to be a Gaussian random variable.
Each type of disorder breaks a different combination of global symmetries:  
$V_{\vec{A}}$ breaks ${\bf P}$ and ${\bf T}$, but preserves ${\bf PH}$; 
$V_M$ breaks all three symmetries;
$V_U$ breaks ${\bf PH}$, but preserves ${\bf P}$ and ${\bf T}$. 

For free massless Dirac fermions (e.g., $L_{\mathrm{QED}_3}$ when $g = 0$), the following is known~\cite{Ludwig1994}: 
(1) $V_{\vec{A}}$ is exactly marginal and results in a fixed line with universal conductivity;
(2) $V_M$ is marginally irrelevant and so the clean fixed point is stable;
(3) $V_U$ is a marginally relevant perturbation.

Our goal is to see how a fluctuating gauge field alters these conclusions.  
Simple power counting shows that the engineering mass dimension of the gauge coupling $\left[ g \right] = 1/2$ and is strongly relevant, leading to the failure of na\"{i}ve perturbation theory.\footnote{This should be contrasted with the problem of graphene with marginal $1/r$ interactions.}  
Nevertheless, the problem is tractable when the number of fermion flavors is large.
In this controlled limit, QED$_3$ flows to an interacting fixed point~\cite{Appelquist:1981sf}. 
In a recent study~\cite{Goswami2017} of the effects of disorder on this fixed point (see also Ref. \cite{Thomson2017}), 
it was found that (1) $V_U$ is strongly irrelevant at the clean interacting fixed point due to the screening of disorder by the longitudinal gauge field fluctuations
and that (2) a $\mathbf{P}$, $\mathbf{T}$, and $\mathbf{PH}$-invariant analogue of $V_M$
leads to a perturbatively accessible finite disorder fixed point corresponding to a dirty metallic phase with universal conductivity.  
While not directly studied in~\cite{Goswami2017}, it is clear that 
$V_{\vec{A}}$ is irrelevant at the clean interacting fixed point which is therefore stable when the number of fermion flavors is large. 

In what follows we will study the effects of disorder in theories consisting of a single ($N= 1$) or two ($N = 2$) copies of $L_{\mathrm{QED}_3}$, where duality arguments are essential.   
The remarkable feature is that the strongly coupled $U(1)^{N}$ QED$_3$ problem becomes dual to $N$ free fermions. 

\section{Duality of free Dirac fermions in the clean limit}
\label{sec:duality}

Duality is a powerful tool for the study of strongly interacting systems.  
In $2+1$ dimensions, many duality transformations are related to the fact that  conserved currents are dual to field strengths of an emergent gauge field:
\begin{equation}
\partial_{\mu} J^{\mu} = 0  \Rightarrow J_{\mu}  = \frac{1}{2 \pi} \epsilon^{\mu \nu \lambda} \partial_{\nu} a_{\lambda}.
\end{equation}
A familiar example, where this occurs, relates bosonic charged particles ({\it e.g.}, Cooper pairs) to vortices in $2+1$ dimensions \cite{Peskin:1977kp, DasguptaHalperin1981}, and plays a crucial role in theories of the superfluid-insulator transition.  
Global currents $J_{\mu}$ associated with the Cooper pair charges  couple to the electromagnetic vector potential $A_{\mu}$, and are expressed in terms of dual electric and magnetic fluxes of $a_{\mu}$.   Vortices, which are neutral with respect to electromagnetism, interact through a logarithmic potential mediated by an emergent gauge field $a_{\mu}$.  
Similar duality transformations also occur in effective descriptions of fractional quantum Hall states \cite{FROHLICH1991517, wen1995}.  Below we will make use of fermionic analogs of such duality transformations. 

Duality is especially useful if it maps a strongly interacting description of a system onto a weakly interacting one.  
The basic building block that we use in our present study is the remarkable statement that $L_{\mathrm{QED}_3}$ is dual to a {\it free} Dirac fermion 
\cite{Son2015, WangSenthilfirst2015, MetlitskiVishwanath2016, Karch:2016sxi,Seiberg:2016gmd, 2016arXiv160601912M}.
We review this duality in Sec.~\ref{subsec:Nf1} and slightly generalize it in Sec.~\ref{subsec:Nf2}.
A more precise description of this duality can be found in Appendix \ref{app:fermion}. 

\subsection{$N=1$ duality}\label{subsec:Nf1}

The $N=1$ duality relates a free Dirac fermion (``theory A'') to QED$_3$ with single fermionic flavor (``theory B''),
\begin{eqnarray}\label{eq:duality1}
L_A [ A ] &=& i\bar\Psi\slashed D_{ A} \Psi - m_{\Psi} \bar \Psi \Psi - \frac{1}{8\pi} AdA + \ldots \nonumber \\
&\updownarrow&  \nonumber \\
L_B[ A ]  &=&  i\bar\psi \slashed D_a\psi - m_{\psi} \bar \psi \psi -\frac{1}{4\pi} Ada - \frac{1}{8\pi} AdA + \ldots
\end{eqnarray}
where $\leftrightarrow$ signifies that the two theories are dual to one another, $A$ is a background non-dynamical gauge field, $AdB \equiv \epsilon_{\mu \nu \lambda} A^{\mu} \partial^{\nu} B^{\lambda}$ represent Chern-Simons terms, and the ellipses denote terms that are irrelevant at low energies. 
For simplicity of presentation, we keep Maxwell terms like $\frac{1}{4g^2} f_{\mu\nu}^2$ for the dynamical gauge field implicit.
Eq. \eqref{eq:duality1} describes an IR duality that is valid for energy scales $E \ll g^2$. 

 Duality maps the global $U(1)$ currents of theory A to electric and magnetic fluxes of theory B,
\be
\label{fermionduality}
J_\mu=\bar \Psi \gamma_\mu \Psi=  -\frac{1}{4 \pi} \epsilon_{\mu \nu \lambda} \partial^{\nu} a^{\lambda},
\ee
and relates the fermion masses near the fixed point,
\be
m_\Psi \sim - m_\psi\,.
\ee
The equivalence between global currents and dual electric and magnetic fluxes in \eqref{fermionduality} also occurs in the conventional bosonic charge-vortex duality \cite{Peskin:1977kp, DasguptaHalperin1981}.  Thus, the field $\Psi$ may rightly be thought of as a fermionic vortex in the theory of $\psi$ and {\it vice-versa}.

The duality above is reviewed more completely in the Appendix \ref{app:fermion}.  
For now, it suffices to note how the same phase diagram can be described on both sides of the duality. 
Recall that when a massive two-component Dirac fermion is integrated out in $2+1$ dimensions, it contributes a level-$1/2$ Chern-Simons term to the effective Lagrangian at long distances:
\begin{equation}
L[b]= i\bar\Psi\slashed D_{ b} \Psi - m_{\Psi} \bar \Psi \Psi \rightarrow L_{eff}[b] = \frac{{\rm sgn} m_{\Psi}}{8 \pi} bdb + \cdots
 \end{equation}
From this, it trivially follows that the two phases of theory A correspond to a gapped insulator ($L_{eff}[A] = 0$) when $m_{\Psi} > 0$, or an integer quantum Hall state ($L_{eff}[A] = -\frac{1}{4 \pi} A dA $) when $m_{\Psi} < 0$.  
Next, consider theory B.  
When the fermions are massive, we can integrate them out (at weak coupling) and then integrate out $a_{\mu}$.  
The latter is done by replacing $a_{\mu}$ by its classical saddle point value.  We then find an integer quantum Hall state  ($L_{eff}[A] = -\frac{1}{4 \pi} A dA$) when $m_{\psi} > 0$, and an insulator ($L_{eff}[A] = 0$) when $m_{\psi} < 0$.  
Thus, the phase diagrams on both sides of the duality match.  

We are interested in gapless metallic states, so let us consider the symmetries of $L_A$ and $L_B$ when $m_\Psi=0$ and $m_\psi=0$.
As is well-known, in the absence of topological order, a single massless 2-component Dirac fermion
can exist in a purely $2+1$ dimensional system only when parity and time-reversal are broken. In our case, ${\bf P}$ and ${\bf T}$ are explicitly broken by the Chern-Simons term in (\ref{eq:duality1}). Since this breaking allows nonzero masses, reaching a gapless state generally requires fine-tuning the fermionic mass to zero.
The duality above can therefore describe a critical point. 
Physically, this critical point arises at the transition between two integer quantum Hall states, whose respective Hall conductances differ by $\Delta \sigma_{xy} = \pm e^2/h = \pm \frac{1}{2 \pi}$.  

Here, we take a different viewpoint. 
The gapless theory is invariant under a ${\bf PH}$ transformation with respect to a filled Landau level.
In theory A, this modified ${\bf PH}$ transformation corresponds to a ${\bf PH}$ transformation followed by a shift of the Lagrangian: $L\mapsto L+\frac{1}{4\pi}AdA$. 
This shift is the modular $\mathcal T$ transformation \cite{WittenSL2Z2003} that physically corresponds to the addition of a filled Landau level \cite{Kivelson1992}. 
When there is a background magnetic field, this transformation implements the familiar mapping $\nu\mapsto1-\nu$, where $\nu$ is the Landau level filling fraction.
In theory B, the ${\bf PH}$ transformation with respect to a filled Landau level includes a ${\bf T}$ transformation on all dynamical fields.
Mass terms are forbidden by this modified ${\bf PH}$ transformation and so no fine-tuning is required in order to reach the gapless state.
Consequently, dual theories invariant under the ${\bf PH}$ transformation with respect to a filled Landau level describe a phase of matter, rather than a critical point. 

\subsection{$N=2$ duality}\label{subsec:Nf2}

We now consider a simple duality that involves two Dirac fermions.
In this case, the gapless state can be ensured without recourse to a modular ${\cal T}$ transformation.

First we take two decoupled copies of the $N = 1$ duality, each with different background gauge fields $A_1, A_2$. This gives a duality between $L_{A}[A_1, A_2] \equiv L_A[ A_1 ] + L_A[ A_2 ]$ and $L_B[ A_1, A_2 ] \equiv L_B [ A_1 ] + L_B [ A_2 ]$.  Note  that $L_B[ A_1, A_2 ]$ will involve two different emergent gauge fields, which we denote $a_1, a_2$. We then introduce the linear combinations  of background gauge fields $A_{\pm} = \left( A_1 \pm A_2 \right)/2$, and similarly for the emergent gauge fields. Finally, applying a $\mathcal T$ transformation to subtract the Chern-Simons terms of $A_\pm$, we arrive at the correspondence:
\begin{eqnarray}\label{eq:duality2}
L_A\left[ A_+, A_- \right] &=&  i \bar \Psi_1 \slashed D_{A_+ + A_-} \Psi_1 + i \bar \Psi_2 \slashed D_{ A_+ - A_-} \Psi_2 - m_{\Psi_j} \bar \Psi_j \Psi_j 
 + \ldots
  \\
&\updownarrow& \nonumber \\
L_B \left[A_+, A_- \right] &=& i \bar \psi_1 \slashed D_{a_+ + a_-} \psi_1 + i \bar \psi_2 \slashed D_{a_+ - a_-} \psi_2  - m_{\psi_j} \bar \psi_j \psi_j
- \frac{1}{2 \pi} \left( a_+ d A_+ + a_- d A_- \right)+\ldots\nonumber
\end{eqnarray}
As before, the terms in `$\ldots$' are irrelevant at long distances.
The mapping of currents and masses follows from the $N=1$ duality,
\be
J^\pm_\mu= i (\bar \Psi_1 \gamma_\mu \Psi_1 \pm \bar \Psi_2 \gamma_\mu \Psi_2)=-\frac{1}{2\pi} \epsilon_{\mu\nu\lambda} \partial^\nu a^\lambda_\pm\;,\; m_{\Psi_j} \sim - m_{\psi_j}\,.
\ee

We are interested in a specific realization of this correspondence that, for instance, gives rise to a dual description for (spinless) graphene. 
In order to obtain it, set the external field $A_-=0$ (recall that these are background fields, which can thus be chosen as we please):
\begin{eqnarray}\label{eq:duality2a}
L_A\left[ A_+,0 \right] &=&  i \bar \Psi_1 \slashed D_{ A_+ } \Psi_1 + i \bar \Psi_2 \slashed D_{ A_+ } \Psi_2 - m_{\Psi_j} \bar \Psi_j \Psi_j 
 + \ldots
 \nonumber \\
&\updownarrow& \nonumber \\
L_B \left[A_+, 0\right] &=& i \bar \psi_1 \slashed D_{a_+ + a_-} \psi_1 + i \bar \psi_2 \slashed D_{a_+ - a_-} \psi_2 - m_{\psi_j} \bar \psi_j \psi_j
- \frac{1}{2 \pi}  a_+ d A_+  + \ldots
\end{eqnarray}
This is a duality between two free Dirac fermions (theory A), a.k.a. spinless graphene with two Dirac points, and a $U(1) \times U(1)$ gauge theory (theory B).\footnote{It will be clear from the context if theories A and B belong to the $N=1$ or $N=2$ duality.} 
Disorder can provide a coupling between the two sets of fermions of theory A or theory B.
We will show how the $N=2$ duality can provide additional non-trivial examples of interacting metallic phases in the presence of disorder.

Let us now discuss the symmetries at the gapless point $m_{\Psi_i}=m_{\psi_i}=0$. 
In addition to an invariance under ${\bf P}$, ${\bf T}$, and ${\bf PH}$, theory A has a $U(1)_+ \times SU(2)$ global symmetry. 
The ``electromagnetic" $U(1)_+$ symmetry rotates $\Psi_j \to e^{i \alpha} \Psi_j$ and its current couples to $A_+$;
$SU(2)$ acts by matrix multiplication on $(\Psi_1, \Psi_2)^{\rm T}$ with the current associated to the $U(1)_-$ subgroup of $SU(2)$ coupling to $A_-$.
These symmetries can be used to enforce the masslessness of the $\Psi_j$ without recourse to a modular ${\cal T}$ transformation.
To see this, it is convenient to parameterize the fermion masses,
\be
m_{\pm}(\bar{\Psi}_1\Psi_1\pm\bar{\Psi}_2\Psi_2) + m_\times \bar{\Psi}_1 \Psi_2 + {\rm h.c.}
\ee
The mass term $m_+$ preserves the $SU(2)$ flavor symmetry while breaking ${\bf P}$, ${\bf T}$, and ${\bf PH}$. 
The $m_-$ and $m_\times$ terms preserve ${\bf P}$, ${\bf T}$, and ${\bf PH}$, but break $SU(2) \to U(1)$.\footnote{For the $N=2$ duality, our convention is that ${\bf T}$ acts on the Dirac fermions by $\Psi_i \mapsto \gamma^2 \Psi_i$ followed by the interchange $\Psi_1 \leftrightarrow \Psi_2$.}

In theory B, only the $U(1)_+ \times U(1)_-$ global symmetries are explicit;
the associated currents are given by the field strengths of $a_\pm$.
A nontrivial prediction of the $N=2$ duality is that theory B has an emergent $SU(2) \supset U(1)_-$ symmetry at low energies.\footnote{Emergent global symmetries are common in duality. 
They occur, for instance, in a supersymmetric mirror symmetry duality \cite{Intriligator:1996ex, Kapustin:1999ha} that has recently been shown to imply the $N=1$ duality \cite{Seiberg:2016gmd, Kachru:2016rui, Kachru:2016aon}.
See Ref.~\cite{2017JHEP...04..135B} for a detailed study of the global symmetries in these and related dualities.}
Duality implies that ${\bf P}$, ${\bf T}$, ${\bf PH}$, and $U(1)_+ \times SU(2)$ symmetries remain unbroken in theory B, if they are preserved in theory A.
Consequently, the preservation of these global symmetries requires both theories to remain gapless (in the clean limit) with theory B describing a strongly coupled phase rather than a critical point.

Once again, let us check that the phase diagrams produced by both Lagrangians match.  Consider first the case when $m_{+} =   0, m_{ -} \neq 0$.  
In this case, integrating out the fermions in theory A, we find an insulator $(L_{eff}[A_{+}] = 0)$.  Viewed from theory B, the two flavors again have opposite sign masses, and after integrating out the fermions, we find, $L_{eff}[A_+] = -\frac{1}{2 \pi} a_+ d( a_- + A_+)$.  We can now safely integrate out the emergent gauge fields.  
Integrating out $a_{-}$ first, we see that we are left with the constraint $\langle a_{+} \rangle = 0$, i.e., the gauge field $a_+$  is ``Higgsed."  
This situation corresponds to an insulator of $J_+$ currents ($L_{eff}[A_{+}] = 0$).  
The description of the insulator of theory A in terms of a superconductor of theory B  is one of the hallmarks of particle-vortex duality \cite{Peskin:1977kp, DasguptaHalperin1981}.

Next, consider the case when $m_{+} \neq 0, m_-=0$.  In this case, the massive fermions of theory A produce an integer quantum Hall effect 
In theory B, both fermions have mass $\mp m_+$.  Integrating out the fermions, we find $L_{eff}[A_+] = -\frac{1}{4 \pi} a_+d a_+ - \frac{1}{4 \pi} a_-da_- - \frac{1}{2 \pi} a_+ dA_+$.  Integrating out $a_{\pm}$, we capture the same effective Lagrangian as before ($L_{eff}[A_+] = \pm \frac{1}{4 \pi} A_+ d A_+$).  
In this case, both sets of fermions are in an integer quantum Hall state.  
\section{Conductivity dictionary}
\label{conductdictionary}

The ``electrical" conductivity defines the diffusive $T=0$ metallic states of interest and provides the primary observable that is sensitive to the effects of disorder.
In this section, we review the dictionary that relates the theory A and theory B expressions for the electrical conductivity.

\subsection{$N=1$}

The background gauge field $A_\mu$ (electromagnetism) couples to the conserved $U(1)$ symmetry.
The $N=1$ duality identifies:
\begin{align}
\bar{\Psi} \gamma_\mu \Psi = - {1 \over 4 \pi} \epsilon_{\mu \nu \rho} \partial^\nu a^\rho.
\end{align}
Consequently, we can relate:
\begin{align}
\label{2pointrelationA}
\langle \bar{\Psi} \gamma_\alpha \Psi(-p) \bar{\Psi} \gamma_\mu \Psi(p) \rangle = {1 \over 16 \pi^2} \epsilon_{\alpha \beta \gamma} \epsilon_{\mu \nu \rho} p^\beta p^\nu \langle a^\gamma(-p) a^\rho(p) \rangle,
\end{align}
where the 3-momenta $p_\mu = (\omega, k_x, k_y)$ and the temperature is left implicit.
Setting $k_x = k_y = 0$ and including the contribution from the background Chern-Simons term for electromagnetism found in \eqref{eq:duality1}, the linear electrical conductivity in theory A,\footnote{More precisely, we can compute \eqref{2pointrelationA} in an imaginary time formulation, which naturally includes temperature effects. \eqref{theoryAconductivity}  is then obtained in the standard way by analytically continuing the frequencies back to the real axis.  
All this is implicit here to preserve clarity.}
\begin{align}
\label{theoryAconductivity}
\sigma_{jk}(\omega) = {\langle \bar{\Psi} \gamma_j \Psi(- \omega) \bar{\Psi}\gamma_k \Psi(\omega) \rangle \over i \omega} - {1 \over 4 \pi} \epsilon_{jk}.
\end{align}
In Lorentz gauge ($\partial_\mu a^\mu = 0$), the emergent gauge field propagator $G_{\mu \nu}$ can be written in terms of its exact self-energy $\pi_{\mu \nu}$
\begin{align}
D_{\mu \nu}^{-1}(p) = - \eta_{\mu \nu} {p^2 \over g^2} + \pi_{\mu \nu}
\end{align}
from which we define the theory B Dirac fermion linear conductivity,
\begin{align}
\sigma_{j k}^B(\omega) = {\pi_{j k}(\omega) + \eta_{j k} \omega^2/g^2 \over i \omega},
\end{align}
corresponding to the current $j_\mu = \delta L_{B}/\delta a^\mu$.
Combining \eqref{2pointrelationA} with \eqref{theoryAconductivity} and taking the IR limit for which $1/g^2 \rightarrow 0$, we find: 
\begin{align}
\label{eq: N=1 dictionary}
\sigma_{i j}(\omega) = {1 \over (4 \pi)^2} \epsilon_{i k} \epsilon_{j l} \rho^{\rm B}_{k l}(\omega) - \frac{1}{4 \pi} \epsilon_{ij}.
\end{align}
where $\rho^{\rm B} = (\sigma^{\rm B})^{-1}$ is the theory B Dirac fermion linear resistivity.

To provide intuition, we describe a concrete realization of the conductivity dictionary.  Consider first the clean limit of theory A, in which the background potential $A_{\mu}$ produces a non-zero magnetic field perpendicular to the plane.  
From \eqref{eq:duality1}, theory A describes fermions in a half-filled Landau level when the chemical potential is tuned to the Dirac point.  
In the clean limit, we can deduce from galilean invariance alone that the conductivity of this system is $\left( \sigma_{xx}, \sigma_{xy} \right) = \left( 0, 1/4\pi \right)$.  
Now consider theory B.  
In the clean limit, the fermions of theory B fill up a Fermi sea and have time-reversal symmetry.  Since momentum relaxation is nearly absent in the clean limit, we expect a diverging longitudinal conductivity and zero Hall conductivity, in accordance with the conductivity dictionary above.  Note that no assumption has been made on existence of quasiparticles or weak-coupling to deduce these results.

\subsection{$N=2$}

The $N=2$ duality identifies the electrical currents of the two theories:
\begin{align}
\bar{\Psi}_n \gamma_\mu \Psi_n = - {1 \over 2 \pi} \epsilon_{\mu \nu \rho} \partial^\nu a^\rho_+,
\end{align}
where $n = 1, 2$.
Identical logic as used for the $N=1$ duality results in a definition of the theory B Dirac fermion linear conductivity in terms of the $a_+$ self-energy.
This self-energy receives corrections from the fluctuations of the $a_-$ gauge field in addition to the other dynamical fields of the theory.
Because there are no background Chern-Simons terms for electromagnetism, the resulting electrical conductivity in the $N=2$ duality\footnote{We have assumed spatial isotropy upon disorder averaging, which requires $\sigma_{xx} = \sigma_{yy}$, and $\sigma_{xy} = - \sigma_{yx}$ for all conductivities.},
\begin{align}
\label{eq: N=2 dictionary}
\sigma_{i j}(\omega) = {1 \over (2 \pi)^2}  \rho^B_{i j} (\omega)
\end{align}

We have seen that for both the $N=1$ and $N=2$ dualities, the conductivity tensor of one set of fields is related to the resistivity tensor of the dual degrees of freedom.  Similar relations also occur  in the bosonic particle-vortex duality.  In that case, when the vortex conductivity vanishes ({\it i.e.}, when the vortices are gapped and form an insulator), the charge degrees of freedom condense into a superfluid.  Conversely, when the vortices  condense into a superfluid, the charges degrees of freedom are gapped, forming an electrical insulator.  In the fermionic analog, the conductivity dictionary says that if the theory A degrees of freedom are localized, the dual theory B Dirac fermions are delocalized, forming a perfect conductor.
Theory B describes a strongly interacting metal when the theory A Dirac fermions are metallic.

\section{Including quenched disorder}
\label{sec:disorder}

We now include the effects of quenched disorder into the $N=1$ and $N=2$ dualities discussed in the previous section.  
To determine the fate of the disordered strongly coupled theory B, we use known results about its free dual, theory A.  
We thereby include the non-perturbative effects of strong gauge interactions in theory B in the presence of quenched disorder.

\subsection{Disorder in the $N=1$ duality}
\label{subsec:disorder1}

\subsubsection{Random flux in theory A $\longleftrightarrow$ Random potential in theory B}\label{subsec:magneticA1}

First consider random magnetic flux in theory A. 
In this case, the spatial components $A_i$ of the background gauge fields are random variables that only depend on space, but not on time.
Theory A is perturbed by
\be
V_{\vec{A}}=\bar\Psi\vec{\gamma}\cdot\vec{A}(\vec{x})\Psi.
\ee
We consider Gaussian white noise disorder with moments,
\be
\overline{ A_j } = 0, \ \ \overline{ A_i(k) A_j(k') } = (2\pi)^2 \delta^2(k+k') \Delta_A\,\delta_{ij}\,,
\ee
where the overline denotes averaging over disorder configurations. 
This disorder therefore gives rise to an external magnetic field $B = \epsilon_{ij} \partial_i A_j$ with zero mean and variance,
\be
\overline{ B(k) B(k') } = (2\pi)^2 \delta^2(k+k') k^2\,\Delta_A\,.
\ee
Note that the magnetic disorder does not contribute a uniform magnetic flux.  

Flux disorder is exactly marginal at the non-interacting fixed point \cite{Ludwig1994}: there is a line of fixed points indexed by $\Delta_A$. 
Along this line, the theory A fermions $\Psi$ acquire a dynamical exponent,
\be\label{eq:dynamicexp}
z= 1+\frac{\Delta_A}{\pi},
\ee
and density of states,
\be\label{eq:density}
\rho(E) \sim \text{Im}(\Psi^\dag \Psi) \sim E^{\frac{\pi-\Delta_A}{\pi+\Delta_A}}\,.
\ee
This fixed line describes a dirty metallic phase with extended (multifractal) wave functions. As one moves along the fixed line, the participation ratio and conductivity vary smoothly.  

The zero-temperature dc conductivity along the fixed line is independent of $\Delta_A$ \cite{Ludwig1994}, however, its value depends upon how the zero temperature and zero frequency limits are approached \cite{damlesachdev97}.
The ``optical conductivity,"
\begin{align}
\sigma_{xx}\Big({T \over \omega} \rightarrow 0\Big) & = {1 \over 16},\cr
\sigma_{xy}\Big({T \over \omega} \rightarrow 0\Big) & = - {1 \over 4 \pi}.
\end{align}
On the other hand, the {\it Landauer} ``dc conductivity,"
\begin{align}
\overline{\sigma}_{xx}\Big({\omega \over T} \rightarrow 0\Big) & = {1 \over \pi}, \cr
\overline{\sigma}_{xy}\Big({\omega \over T} \rightarrow 0\Big) & = - {1 \over 4 \pi}.
\end{align}
The Landauer conductivity is defined by taking $\omega \rightarrow 0$, then $T \rightarrow 0$, at non-zero electron lifetime (imaginary part of the electron self-energy) \cite{PhysRevB.75.205344}.

What do these results tell us about theory B?
The classical action in theory B reads:
\be
L_B= i \bar \psi \slashed D_{a} \psi-\frac{1}{4g^2}f_{\mu\nu}f^{\mu\nu}-\frac{1}{4\pi} a_t B(x) -\frac{1}{8 \pi}  A d A\,.
\ee
Taking the IR limit $g^2 \to \infty$, $a_t$ becomes a Lagrange multiplier that enforces
\be\label{eq:random-density}
\langle \psi^\dag \psi \rangle = \frac{1}{4\pi} B(x)\,.
\ee
Thus, in theory B, random electromagnetic flux results in a disordered {\it density} of theory B fermions.
Since the canonical and grand canonical ensembles are known to be equivalent descriptions in the thermodynamic limit, we expect that the problem being solved in theory B can equivalently be cast as that of a random chemical potential, determined self-consistently from \eqref{eq:random-density}.\footnote{Clearly, there must be some quenched random chemical potential that produces the random density pattern.}

We may extract optical and dc conductivities for the Dirac fermions of theory B using the dictionary in \eqref{eq: N=1 dictionary}. 
We find the theory B Dirac fermion ``optical conductivity,''
\begin{align}
\sigma^B_{xx}\Big({T \over \omega} \rightarrow 0\Big) & = {1 \over \pi^2},\cr
\sigma^B_{xy}\Big({T \over \omega} \rightarrow 0\Big) & = 0.
\end{align}
On the other hand, the theory B Dirac fermion Landauer ``dc conductivity,"
\begin{align}
\overline{\sigma}^B_{xx}\Big({\omega \over T} \rightarrow 0\Big) & = {1 \over 16 \pi}, \cr
\overline{\sigma}^B_{xy}\Big({\omega \over T} \rightarrow 0\Big) & = 0.
\end{align}
Using the $N=1$ duality, we conclude that QED$_3$ in the presence of a random chemical potential (determined self-consistently from \eqref{eq:random-density}) does not localize.
Instead, there is a line of dirty metals parameterized by $\Delta _A$. 
Since the vortices of theory B correspond to the free Dirac fermions of theory A, we learn that the vortices of QED$_3$ are characterized by a dynamical exponent $z$ given by (\ref{eq:dynamicexp}), have a multifractal wavefunction, and acquire a density of states (\ref{eq:density}). 
It would be very interesting to test these predictions numerically.

Let us compare this result with different perturbative analyses. First, for a free Dirac fermion (the approximation of weak coupling in theory B), random chemical potential disorder is marginally relevant. 
On the other hand, Ref.~\cite{Goswami2017} showed that chemical potential disorder is irrelevant in an approximation where the number of fermion flavors is large. 
The reason is that at the clean large flavor fixed point, the electric part of the gauge fluctuations screen the disorder potential, making it short-range correlated. 
Our approach based on duality allows us to access the QED$_3$ theory with a single flavor at strong coupling and reveals the existence of a fixed line for arbitrary disorder variance. 
One might speculate that the screening physics of the large $N$ theory counterbalances the quantum corrections that cause the disorder to grow in the free theory.

\subsubsection{Random potential in theory A $\longleftrightarrow$ Random flux in theory B}\label{subsec:chemicalA1}

Next consider a random chemical potential in theory A, which is obtained from disordering $A_t \equiv U$,
\be
V_{U} = U(\vec{x}) \Psi^\dagger \Psi.
\ee
For Gaussian white noise disorder,
\be
\overline{U(k) } = 0, \ \ \overline{ U(k) U(k') }= (2\pi)^2 \delta^2(k+k') \Delta_U\,,
\ee
a weak random chemical potential is marginally relevant at the non-interacting fixed point~\cite{Ludwig1994}.  
Unbroken {\bf PH} symmetry with respect to a filled Landau level prevents the localization of the theory A fermion.

In contrast to the case of flux disorder, the diffusive fixed point obtained from a random chemical potential in theory A has finite and equal optical and dc conductivities \cite{Huo1993}:
\begin{align}
\sigma_{xx}\Big({\omega \over T} \rightarrow 0\Big) & = \sigma_{xx}\Big({T \over \omega} \rightarrow 0\Big) \approx {1 \over 4 \pi},\cr
\sigma_{xy}\Big({\omega \over T} \rightarrow 0\Big) & = \sigma_{xy}\Big({T \over \omega} \rightarrow 0\Big) = - {1 \over 4 \pi}.
\end{align}

Duality tells us that random chemical potential in theory A manifests itself in theory B as a random vector potential.
This can be seen from the coupling between $A_{\mu} $ and the emergent gauge fields $a_{\mu}$:
\be
V_U = -\frac{1}{4\pi}U(\vec{x}) \,\epsilon_{ij} \partial_i a_j\,,
\ee
that sources the random emergent magnetic field in theory B.

Duality implies that random flux is relevant in QED$_3$. 
Using the conductivity dictionary \eqref{eq: N=1 dictionary}, the theory B Dirac fermion conductivity,
\begin{align}
\sigma^B_{xx}\Big({\omega \over T} \rightarrow 0\Big) & = \sigma^B_{xx}\Big({T \over \omega} \rightarrow 0\Big) \approx {1 \over 4 \pi},\cr
\sigma^B_{xy}\Big({\omega \over T} \rightarrow 0\Big) & = \sigma^B_{xy}\Big({T \over \omega} \rightarrow 0\Big) = 0.
\end{align}
Thus, random flux in theory B results in a diffusive metallic fixed point.

\subsubsection{Random mass in theory A $\longleftrightarrow$ Random mass in theory B}\label{subsec:massA}
\label{subsub: N=1 random mass}
Finally, we disorder theory A by a random mass,
\be
V_M=M(\vec{x})\bar\Psi\Psi.
\ee
Such disorder is marginally irrelevant: the clean fixed point is stable.

In Appendix \ref{app:freeDirac}, we show that the optical conductivity,
\begin{align}
\sigma_{xx}\Big({T \over \omega} \rightarrow 0\Big) & = {1 \over 16},\cr
\sigma_{xy}\Big({T \over \omega} \rightarrow 0\Big) & = - {1 \over 4 \pi}.
\end{align}
By the same calculation, the dc conductivity,
\begin{align}
\sigma_{xx}\Big({\omega \over T} \rightarrow 0\Big) & = 0, \cr
\sigma_{xy}\Big({\omega \over T} \rightarrow 0\Big) & = - {1 \over 4 \pi}.
\end{align}
See Ref.~\cite{PhysRevB.75.205344} for discussion on the sensitivity of the conductivity to the zero temperature and frequency limit for a free Dirac fermion without disorder.

A random mass in theory A corresponds to a random mass in theory B.
Consequently, we learn that random mass disorder in theory B is marginally irrelevant.
The conductivity dictionary allows us to immediately infer the theory B Dirac fermion response.
The theory B Dirac fermion optical conductivity,
\begin{align}
\sigma^B_{xx}\Big({T \over \omega} \rightarrow 0\Big) & = {1 \over \pi^2},\cr
\sigma^B_{xy}\Big({T \over \omega} \rightarrow 0\Big) & = 0;
\end{align}
the theory B Dirac fermion dc resistivity,
\begin{align}
\rho^B_{xx}\Big({\omega \over T} \rightarrow 0\Big) & = 0, \cr
\rho^B_{xy}\Big({\omega \over T} \rightarrow 0\Big) & = 0.
\end{align}
Thus, the theory B Dirac fermions  realizes a perfect conductor, a state with zero resistance.
 
\subsection{Disorder in the $N=2$ duality}
\label{subsec:disorder2}

The results for the $N=1$ duality readily carry over to the $N=2$ duality, albeit with the modified conductivity dictionary Eq. \eqref{eq: N=2 dictionary}.
If the disorder is chosen to preserve the $SU(2)$ symmetry, the $N=2$ duals behave as two decoupled copies of the $N=1$ duals.

\subsubsection{Random flux in theory A $\longleftrightarrow$ Random potential in theory B}\label{subsec:magneticA2}

Flux disorder in theory A corresponds to the term,
\be
V_{\vec{A}} = \bar \Psi_n \vec \gamma\cdot \vec{A}(\vec{x}) \Psi_n,
\ee
where $n = 1 ,2$.
As in Sec.~\ref{subsec:magneticA1}, this perturbation is exactly marginal and results in a line of diffusive interacting metals, using the $N=2$ duality, that are parameterized by the disorder variance $\Delta_A$. 

The electrical optical conductivity,
\begin{align}
\sigma_{xx}\Big({T \over \omega} \rightarrow 0\Big) & = {1 \over 8},\cr
\sigma_{xy}\Big({T \over \omega} \rightarrow 0\Big) & = 0.
\end{align}
On the other hand, the {\it average} electrical dc conductivity
\begin{align}
\overline{\sigma}_{xx}\Big({\omega \over T} \rightarrow 0\Big) & = {2 \over \pi}, \cr
\overline{\sigma}_{xy}\Big({\omega \over T} \rightarrow 0\Big) & = 0.
\end{align}
Using the $N=2$ conductivity dictionary \eqref{eq: N=2 dictionary}, the optical conductivity of the theory B fermions,
\begin{align}
\sigma^B_{xx}\Big({T \over \omega} \rightarrow 0\Big) & = {2 \over \pi^2},\cr
\sigma_{xy}\Big({T \over \omega} \rightarrow 0\Big) & = 0;
\end{align}
the theory B Dirac fermion {\it average} dc conductivity,
\begin{align}
\overline{\sigma}^B_{xx}\Big({\omega \over T} \rightarrow 0\Big) & = {1 \over 8 \pi}, \cr
\overline{\sigma}^B_{xy}\Big({\omega \over T} \rightarrow 0\Big) & = 0.
\end{align}
In contrast to the $N = 1$ duality, the theory B fermions generically realize a metallic critical point because $V_{\vec{A}}$ breaks {\bf T}.

\subsubsection{Random  potential in theory A $\longleftrightarrow$ Random flux in theory B}\label{subsec:chemicalA2}

Next, consider a random chemical potential in theory A,
\be
\label{chemnequal2}
V_{U} = U(\vec{x}) \Psi_n^\dagger \Psi_n,
\ee
where $n = 1,2$.
As in the $N=1$ case, $V_U$ is a marginally relevant perturbation.
Because the chemical potential disorder in \eqref{chemnequal2} preserves the $SU(2)$ symmetry, the theory again flows to a diffusive fixed point (localization only occurs if additional random perturbations are included).
Thus, we essentially obtain two decoupled copies of the diffusive fixed point of Section \ref{subsec:chemicalA1}:
\begin{align}
\sigma_{xx}\Big({\omega \over T} \rightarrow 0\Big) & = \sigma_{xx}\Big({T \over \omega} \rightarrow 0\Big) \approx {1 \over 2 \pi},\cr
\sigma_{xy}\Big({\omega \over T} \rightarrow 0\Big) & = \sigma_{xy}\Big({T \over \omega} \rightarrow 0\Big) = 0.
\end{align}
Using the conductivity dictionary, the theory B Dirac fermion conductivity,
\begin{align}
\sigma^B_{xx}\Big({\omega \over T} \rightarrow 0\Big) & = \sigma^B_{xx}\Big({T \over \omega} \rightarrow 0\Big) \approx {1 \over 2 \pi},\cr
\sigma^B_{xy}\Big({\omega \over T} \rightarrow 0\Big) & = \sigma^B_{xy}\Big({T \over \omega} \rightarrow 0\Big) = 0.
\end{align}
Because the random flux in theory B preserves the discrete global symmetries of the model that exclude mass terms for the theory B fermions, we obtain a diffusive metallic phase.

\subsubsection{Random mass in theory A $\longleftrightarrow$ Random mass in theory B}\label{subsec:massA2}

Similarly to the $N=1$ duality, random $SU(2)$ preserving mass in theory A and theory B is an irrelevant perturbation to the clean fixed point.
Appendix \ref{app:freeDirac} and the conductivity dictionary enable us to relate the electrical conductivity to the theory B Dirac fermion conductivity.
The electrical optical conductivity,
\begin{align}
\sigma_{xx}\Big({T \over \omega} \rightarrow 0\Big) & = {1 \over 8},\cr
\sigma_{xy}\Big({T \over \omega} \rightarrow 0\Big) & = 0,
\end{align}
while the dc conductivity,
\begin{align}
\sigma_{xx}\Big({\omega \over T} \rightarrow 0\Big) & = 0, \cr
\sigma_{xy}\Big({\omega \over T} \rightarrow 0\Big) & = 0.
\end{align}
The theory B Dirac fermion optical conductivity,
\begin{align}
\sigma^B_{xx}\Big({T \over \omega} \rightarrow 0\Big) & = {2 \over \pi^2},\cr
\sigma^B_{xy}\Big({T \over \omega} \rightarrow 0\Big) & = 0;
\end{align}
the theory B Dirac fermion dc resistivity,
\begin{align}
\rho^B_{xx}\Big({\omega \over T} \rightarrow 0\Big) & = 0, \cr
\rho^B_{xy}\Big({\omega \over T} \rightarrow 0\Big) & = 0.
\end{align}
Again, the theory B Dirac fermions realize a perfect  conducting phase.

\subsubsection{Effect of flavor mixing perturbations}

So far, we have considered cases where the disorder preserves the $SU(2)$ global symmetry of the theory.  
Not surprisingly, by requiring this symmetry to be unbroken, the disorder problem, considered within the $N=2$ duality, is trivially the same as that of two decoupled $N=1$ duals.  
To obtain qualitatively new behavior, we must break the $SU(2)$ symmetry.   

As an illustrative example, consider the case of flux disorder.  
To be concrete, imagine a lattice realization, such as graphene without spins, in which electrons are exposed to a random vector potential on each lattice link. 
In the interest of generality, we do not assume a sublattice symmetry.\footnote{In graphene, nearest neighbor hopping preserves a bipartite sublattice symmetry.  This is no longer true when further neighbor hoppings (which are always present, albeit small) occur.}
In the clean limit, this system has two Dirac nodes (``valleys"), at momenta $\vec k_{1,2}$ in the Brillouin zone.   
If the disorder is predominantly of long-wavelength in character, then its primary effect is to couple fermions within the same valley, i.e., it preserves the $SU(2)$ global symmetry.
Large momentum components of the flux disorder, however, mix the valleys.  
Consequently, the Lagrangian $L_A[A_+, 0]$ obtains a correction of the form,
\begin{align}
\delta L_A[A_+,0] = \bar \Psi_1  \mathcal M_{12} \Psi_2 +
 \Psi_1^{\dagger} \slashed{\cal A}_{12} \Psi_2 + \Psi_1^{\dagger} \mathcal V_{12} \Psi_2,
\end{align}
where $\mathcal M_{12}$,  ${\cal A}_{12}$,  and $\mathcal V_{12}$ are inter-valley mass, ``nonabelian" vector, and scalar potential disorder -- now allowed by symmetry \cite{Chou2014}.  
All three quantities consist of quenched random variables with zero disorder average.     We will think of the inter-valley operators as a perturbation (justified by postulating an arbitrarily smooth disorder potential).
From duality, the problem maps onto that of $SU(2)$-preserving potential disorder in theory B, with perturbative mixing of the two flavors.
While a precise map of the $SU(2)$ mixing terms does not exist, in the spirit of effective field theory, we expect mass, scalar, and vector potential disorder potentials that couple the theory B fermions to be generated by the interactions.

As is well-known, with these additional perturbations, theory A is a member of the unitary symmetry class \cite{Mirlin2008}.  
There are no topological terms present in the effective treatment of the disorder problem, and the system undergoes Anderson localization.  
Thus, the electrical conductivity tensor will be
\begin{align}
\sigma_{xx}\Big({\omega \over T} \rightarrow 0\Big) & = \sigma_{xx}\Big({T \over \omega} \rightarrow 0\Big) = 0,\cr
\sigma_{xy}\Big({\omega \over T} \rightarrow 0\Big) & = \sigma_{xy}\Big({T \over \omega} \rightarrow 0\Big) = 0.
\end{align}
As a consequence, the conductivity dictionary implies the fermions of theory B realize a perfect conductor with vanishing resistivity:     
\begin{align}
\rho^B_{xx}\Big({\omega \over T} \rightarrow 0\Big) & = \rho^B_{xx}\Big({T \over \omega} \rightarrow 0\Big) = 0,\cr
\rho^B_{xy}\Big({\omega \over T} \rightarrow 0\Big) & = \rho^B_{xy}\Big({T \over \omega} \rightarrow 0\Big) = 0.
\end{align}
The reason for this behavior is that the vortices of theory B, i.e., the $\Psi$ fermions,
localize.
We are unable to determine from this analysis whether the theory B Dirac fermions realize a perfect metal \cite{Chakravarty1998, Plamadeala2014} or a superconductor.
Only the latter exhibits a Meissner effect with respect to the emergent $a_+$ gauge field;
the former is expected to exhibit finite conductivity for any $T>0$ or $\omega > 0$.
 

As a second example, consider the case of potential disorder in theory A, again breaking the $SU(2)$ ``valley" symmetry.  
In this case, theory A enjoys both {\bf P} and {\bf T} symmetries, but {\bf PH} is broken and so all components of $\mathcal M_{12}, \mathcal A_{12}, \mathcal V_{12}$ that are consistent with these symmetries are allowed.
For the case of a smooth disorder potential that scatters mainly in the long-wavelength limit within a valley, such valley mixing terms can again be viewed as a perturbation.  
In the language of theory B, the problem is that of $SU(2)$-preserving magnetic flux disorder, with additional $SU(2)$-breaking mass, vector, and scalar potentials that mix the valleys.  
In this case, the fermions of theory A realize the orthogonal symmetry Wigner-Dyson class \cite{Mirlin2008} and undergo Anderson localization. 
Thus, the theory B fermions again realize a  perfect conductor.    

\section{Discussion}\label{sec:concl}

The examples discussed above show that gauge fluctuations can stabilize metallic behavior in two dimensional systems, similar to the conclusions of a recent perturbative study~\cite{Goswami2017}.  The examples studied here all involve Dirac fermions with the chemical potential at neutrality, and hence are all somewhat special.  In particular, the neutrality implies that the current operator has zero overlap with the momentum operator.  Consequently, conductivities may be finite even in the clean limit.  In the future, it will be of great interest to consider systems with Fermi surfaces in the clean limit.  The dualities employed in the present paper can easily be used to study such problems.  It will also be interesting to apply bosonization dualities to study disorder effects on quantum Hall systems.  We will present progress in these directions in future publications. 

The reader may naturally question the relevance of the above results to condensed matter systems.  The strongly coupled examples here describe fermions that do not couple directly to electromagnetism but instead to emergent gauge fields.  Thus, the predictions made here cannot be directly used to explain metallic behavior seen in experiments.  Instead, they should be viewed as an in principle demonstration that such behavior can exist because of strong interactions.  A more direct application of these results would be to certain classes of spin liquids, where spinon degrees of freedom are Dirac fermions and coupled to emergent $U(1)$ gauge fields \cite{Rantner2001,Hermele2005}.  
While unambiguous experimental evidence of such spin liquids is currently lacking, there is much reason to be optimistic.  

Finally, we comment on the nature of the effective field theory of the disorder problem in the presence of strong gauge fluctuations.  
The standard approach for non-interacting systems, is to study the low energy diffusive modes, which are described by a non-linear sigma model, whose coupling corresponds to the (charge) diffusion constant.  
In the  present context, such effective theories would apply to gauge theories when the mean-free path $\lambda \gg 1/g^2$, which is required for duality to hold. 
From duality, we learn that it is the diffusion of flux rather than charge that acts as the coupling constant in the effective theory for the diffusive modes.  
When quantum diffusion of flux is destroyed by disorder, the conductivity diverges.  It is intriguing to ask whether this lesson may apply to a broader class of systems than the ones studied here.  

\section*{Acknowledgments}
We are especially grateful to M. Foster for helpful correspondence and insightful conversations.
We also thank E. Abrahams, S. Chakravarty, T. Faulkner, E. Fradkin, S. Kachru, A. Ludwig, H. Wang, and Y. Wang for discussions.
M.M. was supported in part by the UCR Academic Senate.
M.M. is grateful for the generous hospitality of the Aspen Center for Physics, which is supported by the National Science Foundation (NSF) grant PHY-1607611, and the Kavli Institute for Theoretical Physics, under Grant No. NSF PHY-1125915, where some of this work was performed. H. G. is supported by the NSF under Grant No. DMR-1725401 and by the NSF Graduate Research Fellowship Program under Grant No. DGE-1144245.   S. R.  is supported by the DOE Office of Basic Energy Sciences, contract DE-AC02-76SF00515.  G. T. is supported by CONICET, PIP grant 11220150100299, by ANPCYT PICT grant 2015-1224, and by CNEA.

\appendix
\section{$N=1$ duality}\label{app:fermion}

In this appendix, we review the derivation in \cite{Seiberg:2016gmd} of the $N=1$ duality \cite{Son2015, WangSenthilfirst2015, MetlitskiVishwanath2016, Karch:2016sxi,Seiberg:2016gmd, 2016arXiv160601912M}; 
we follow the conventions of \cite{Kachru:2016rui, Kachru:2016aon}.
Recent studies of the two bosonization dualities listed below include \cite{2017arXiv170505841C, 2017arXiv170501106M}; older works include \cite{ChenFisherWu1993, wenwu1993}.

We begin from the following two dualities, established in \cite{Kachru:2016rui, Kachru:2016aon} using mirror symmetry \cite{Intriligator:1996ex}. 
First, we have a duality between a free Dirac fermion and scalar QED:
\begin{align}\label{eq:pair1}
& i \bar \Psi\slashed D_{A}\Psi-m_\Psi \bar \Psi \Psi -\frac{1}{8\pi}A d A \cr
& \updownarrow \cr
& |D_{-a}u|^2-m_u^2 |u|^2-|u|^4+\frac{1}{4\pi} a da -\frac{1}{2\pi} a dA 
\end{align}
where $A$ is the background gauge field for the $U(1)$ global symmetry, $a$ is a dynamical $2+1$ dimensional gauge field, $D_{\pm a} = \partial_\mu \mp i a_\mu$ with $\mu \in \{t,x,y \}$, $\slashed D_A = \gamma^\mu (\partial_\mu - i A_\mu)$, and $\gamma^\mu = (\sigma^3, i \sigma^1, i \sigma^2)$ where $\sigma^j$ are Pauli matrices. 
Chern-Simons terms are denoted by $AdB = \epsilon^{\mu \nu \rho} A_\mu \partial_\nu B_\rho$.
Close to the phase transition, where all matter fields are gapless, the masses are related by
\be\label{eq:mass1}
m_u^2 \sim -m_\Psi\,.
\ee
Second, we have a duality between Wilson-Fisher and fermion QED,
\begin{align}
\label{eq:pair2}
&|D_{A} v|^2-m_v^2 |v|^2-|v|^4-\frac{1}{4\pi}A d A \cr
& \updownarrow \cr
&i \bar \psi \slashed D_{-a} \psi-m_\psi \bar \psi \psi+\frac{1}{8\pi} a da-\frac{1}{2\pi} a d A.
\end{align}
Again, close to the transition,
\be\label{eq:mass2}
m_v^2\sim m_\psi\,.
\ee

In order to derive the $N=1$ duality, we need to transform the Wilson-Fisher side of (\ref{eq:pair2}) into the scalar QED theory in (\ref{eq:pair1}). 
For this, we add $\frac{2}{4\pi} A dA$ to both sides of (\ref{eq:pair2}) (namely we ${\cal T}$-transform twice), promote $A \to b$ do a dynamical field, and couple it to an external $B$ (i.e., the ${\cal S}$-transform \cite{WittenSL2Z2003}). 
The resulting dual pair is
\begin{align}
\label{eq:pair3}
& |D_b v|^2-m_v^2 |v|^2-|v|^4+\frac{1}{4\pi} b db-\frac{1}{2\pi} b dB \cr
& \updownarrow \cr
& i \bar \psi \slashed D_{-a} \psi-m_\psi \bar \psi \psi+\frac{1}{8\pi}a da-\frac{1}{2\pi} a db+\frac{1}{2\pi}  b db -\frac{1}{2\pi} b d B\,.
\end{align}
After renaming $b \to -b$, $B \to A$, the scalar side of (\ref{eq:pair3}) is the same as that in (\ref{eq:pair1}). Hence we obtain the $N=1$ duality (after also changing $a \to -a$)
\begin{align}
\label{eq:pair4}
L_A[A] =& i \bar \Psi\slashed D_{A}\Psi-m_\Psi \bar \Psi \Psi -\frac{1}{8\pi}A d A \cr
& \updownarrow \cr
L_B[A] =&i \bar \psi \slashed D_{a} \psi-m_\psi \bar \psi \psi+\frac{1}{8\pi}a da-\frac{1}{2\pi} a db+\frac{1}{2\pi}  b db -\frac{1}{2\pi} b d A.
\end{align}
Using the previous relations between mass parameters we find:
\be\label{eq:masses1}
m_\Psi \sim - m_\psi
\ee
close to the transition.

Integrating out $b$ sets $b=\frac{1}{2}(a+A)$ and results in the simplified form of the $N=1$ duality used throughout this paper:
\begin{align}
\label{eq:Son}
L_A[A]=& i \bar \Psi\slashed D_{A}\Psi-m_\Psi \bar \Psi \Psi -\frac{1}{8\pi}A d A\cr
& \updownarrow \cr
L_B[A]=&i \bar \psi \slashed D_{a} \psi-m_\psi \bar \psi \psi-\frac{1}{4\pi} a d A-\frac{1}{8 \pi}  A d A\,.
\end{align}
Maxwell terms $\frac{1}{4g^2} f_{\mu\nu}^2$ for the dynamical gauge fields are implicitly understood throughout this work. 
The duality between $L_A$ and $L_B$ is valid in the IR where $E/g^2  \to 0$, for all relevant energy scales $E$.

\section{Free Dirac fermion conductivity at finite temperature and frequency}\label{app:freeDirac}

In this appendix, we calculate the finite-temperature ac conductivity of a free Dirac fermion.
The Hamiltonian,
\begin{equation}
\hat H(k) = k_x \sigma^x + k_y \sigma^y + m \sigma^z.
\end{equation}
Note that our choice of $\gamma^\mu$ matrices in this appendix differs from that in the main text.
In imaginary time, the fermion propagator is 
\begin{equation}
G(k,i \omega_n) = \left( i \omega_n \hat 1 - \hat H(k) \right)^{-1}  = \sum_{s = \pm 1} \frac{ \hat P_s}{i \omega_n - E_s(k)}, \ \ E_s = s \sqrt{k^2 + m^2} , \ \ \hat P_s = \frac{1}{2} \left( \hat 1 + \frac{ \hat H(k)}{E_s(k)} \right).
\end{equation}
Therefore, the current-current correlation function is 
\begin{eqnarray}
\Pi_{ij} \left( i \Omega_n \right) &=& \frac{1}{\beta} \sum_{s,t=\pm 1} \int \frac{ d^2 k}{(2 \pi)^2} \sum_m {\rm Tr} \left[  \frac{ \sigma_i \hat P_s \sigma_j \hat P_t }{\left(i\omega_m + i \Omega_n - E_s(k) \right) \left( i \omega_m - E_t(k) \right) } \right] ,  \nonumber \\
&=& \sum_{s,t=\pm 1} \int \frac{ d^2 k}{(2 \pi)^2} {\rm Tr} \left[  \sigma_i \hat P_s \sigma_j \hat P_t \right] \frac{ f( t \sqrt{k^2 + m^2}) - f(s  \sqrt{k^2 + m^2})}{i \Omega_n - \left(t-s \right) \sqrt{k^2 + m^2} }, 
\end{eqnarray}
where $\beta = 1/T$ and $f(x) = \frac{1}{ e^{\beta x } + 1}$.
The conductivity is obtained by analytically continuing, $i \Omega_n \rightarrow \omega + i \delta$, the expression,
\begin{equation}
\sigma_{ij}(\omega, T) = \frac{i}{\omega} \Pi_{ij}(\omega + i \delta).
\end{equation}
After computing the traces and using rotational invariance, 
\begin{align} 
\sigma_{xx}(\omega, T) =& \frac{i}{\omega} \int  \frac{ d^2 k}{(2 \pi)^2} \frac{k^2 + 2 m^2}{2 \left( k^2 + m^2 \right)} \left[  \frac{ 1- 2f(  \sqrt{k^2 + m^2})}{\omega - 2 \sqrt{k^2 + m^2} + i \delta} + \frac{2 f(  \sqrt{k^2 + m^2}) - 1}{\omega + 2 \sqrt{k^2 + m^2} + i \delta} \right], \cr
\sigma_{xy}(\omega, T) =& \frac{i}{\omega} \int  \frac{ d^2 k}{(2 \pi)^2} \left[ \frac{-i m}{ \sqrt{k^2 + m^2} }\right]  \left[  \frac{ 1- 2f(  \sqrt{k^2 + m^2})}{\omega - 2 \sqrt{k^2 + m^2} + i \delta} - \frac{2 f(  \sqrt{k^2 + m^2}) - 1}{\omega + 2 \sqrt{k^2 + m^2} + i \delta} \right].
\end{align}
The real part of $\sigma_{xx}$ is obtained by extracting the Dirac delta piece:
\begin{align}
{\rm Re} \left[ \sigma_{xx}  (\omega, T) \right] =& \frac{\pi}{ \omega} \int  \frac{ d^2 k}{(2 \pi)^2} \frac{k^2+ 2 m^2}{2 \left( k^2 + m^2 \right)} \left[ 1 - 2 f(\sqrt{k^2 + m^2} ) \right] \cr
& \times \left[ \delta \left(\omega - 2 \sqrt{k^2 + m^2} \right) - \delta \left(\omega + 2 \sqrt{k^2 + m^2} \right) \right]  \cr
=& \frac{\pi}{2 \omega} \left( 1 +  \frac{m^2}{\omega^2} \right) \tanh{\left[ \frac{\beta \omega}{4} \right]} \cr
\times & \int  \frac{ d^2 k}{(2 \pi)^2} \Big[ \delta \left(\omega - 2 \sqrt{k^2 + m^2} \right) +  \delta \Big(\omega + 2 \sqrt{k^2 + m^2} \Big) \Big]
\end{align}
using
\begin{eqnarray}
\int \frac{ d^2 k}{(2 \pi)^2}  \delta \left(\omega - 2 \sqrt{k^2 + m^2} \right) &=& \frac{1}{2 \pi} \int d k \ k  \frac{\omega}{2} \frac{\delta(k-k_0)}{2 k_0}, \ \ \ k_0 = \sqrt{(\omega/2)^2 - m^2} \nonumber \\
&=&  \frac{\omega}{8 \pi} \Theta( \omega^2 - (2 m )^2).
\end{eqnarray}
Thus, we find:
\begin{equation}
{\rm Re} \left[ \sigma_{xx}  (\omega, T) \right] = \frac{1}{16} \left( 1 +  \frac{m^2}{\omega^2} \right) \tanh{\left[ \frac{\beta \omega}{4} \right]}  \Theta( \omega^2 - (2 m )^2).
\end{equation}
When $m=0$, this expression simplifies to
\begin{equation}
{\rm Re} \left[ \sigma_{xx}  (\omega, T) \right]  = \frac{1}{16}  \tanh{\left[ \frac{\beta \omega}{4} \right]}.
\end{equation}

In a similar way, we can obtain the imaginary part of $\sigma_{xy}$ by extracting the Dirac delta function piece:
\begin{eqnarray}
{\rm Im} \left[ \sigma_{xy}(\omega, T) \right] &=& - \frac{2 \pi m}{\omega^2}  \tanh{\left[ \frac{\beta \omega}{4} \right]} \int  \frac{ d^2 k}{(2 \pi)^2}\delta \left(\omega - 2 \sqrt{k^2 + m^2} \right) \nonumber \\
&=&-\frac{m}{4 \omega} \tanh{\left[ \frac{\beta \omega}{4} \right]} \Theta( \omega^2 - (2 m )^2).
\end{eqnarray}
From the Kramers-Kronig relations,
\begin{eqnarray}
{\rm Re} \left[ \sigma_{xy}(\omega, T) \right] &=& \frac{2}{ \pi} \mathcal P \int_0^{\infty} dx 
\frac{ x  }{x^2 - \omega^2} {\rm Re }
\left[ \sigma_{xy}(x,T) \right] \nonumber \\
&=& -\frac{m}{2 \pi} \mathcal P \int_{2 \vert m \vert}^{\infty} dx \frac{ \tanh{\left[ \frac{\beta x}{4} \right]}}{x^2-\omega^2}.
\end{eqnarray}
The dc result is obtained by setting $\omega = 0$ at finite $T$: 
\begin{eqnarray}
{\rm Re} \left[ \sigma_{xy}(0, T) \right] &=& -\frac{m}{2 \pi} \int_{2 \vert m \vert}^{\infty} dx  \frac{ \tanh{\left[ \frac{\beta x}{4} \right]}}{x^2} \nonumber \\
&\simeq& -\frac{1}{ 4\pi} \frac{m}{\vert m\vert}, \ \ \beta m \gg 1.
\end{eqnarray}
The ac result is obtained by setting $T=0$ at finite $\omega$:
\begin{eqnarray}
{\rm Re} \left[ \sigma_{xy}(\omega, 0) \right] &=& -\frac{m}{2 \pi} \int_{2 \vert m \vert}^{\infty} \frac{dx}{x^2 - \omega^2} ,\ \ \omega < m \nonumber \\
&=& \frac{m}{4 \pi \omega} \log{\left[ \frac{2 \vert m \vert - \omega}{2 \vert m \vert + \omega}  \right] }.
\end{eqnarray}
In this case, the ac result smoothly meets the dc value in the zero frequency limit.  

\bibliography{metal.bib}{}
\bibliographystyle{utphys}

\end{document}